\documentclass[UKenglish]{lipics-v2021}
\pdfoutput=1
\usepackage{graphicx} 
\usepackage[dvipsnames]{xcolor}
\usepackage{amsmath}
\usepackage{dsfont}
\usepackage{bussproofs}
\usepackage{semantic}
\usepackage{tikz-cd}
\usepackage{stmaryrd}
\usepackage{newunicodechar}
\usepackage{mathpartir}
\usepackage{qcircuit}

\nolinenumbers

\DeclareMathAlphabet{\mymathds}{U}{BOONDOX-ds}{m}{n}
\newcommand\llbrace{\mathopen{{\{\kern-0.24em|}}}    
\newcommand\rrbrace{\mathclose{{|\kern-0.24em\}}}}   
\newcommand{\bidirmultimap}{\mathrel{\ooalign{$\mkern6mu\multimap$\cr$\reflectbox{\(\multimap\)}$\cr}}}
\newunicodechar{ }{{\,}}
\newunicodechar{¬}{\ensuremath{\neg}} %
\newunicodechar{²}{^2}
\newunicodechar{³}{^3}
\newunicodechar{·}{\ensuremath{\cdot}}
\newunicodechar{¹}{^1}
\newunicodechar{×}{\ensuremath{\times}} %
\newunicodechar{÷}{\ensuremath{\div}} %
\newunicodechar{ʳ}{^r}
\newunicodechar{ˡ}{^l}
\newunicodechar{Γ}{\Gamma}   %
\newunicodechar{Δ}{\ensuremath{\Delta}} %
\newunicodechar{Η}{\ensuremath{\textrm{H}}} %
\newunicodechar{Θ}{\ensuremath{\Theta}} %
\newunicodechar{Λ}{\ensuremath{\Lambda}} %
\newunicodechar{Ξ}{\ensuremath{\Xi}} %
\newunicodechar{Π}{\ensuremath{\Pi}}   %
\newunicodechar{Σ}{\ensuremath{\Sigma}} %
\newunicodechar{Φ}{\ensuremath{\Phi}} %
\newunicodechar{Ψ}{\ensuremath{\Psi}} %
\newunicodechar{Ω}{\ensuremath{\Omega}} %
\newunicodechar{α}{\ensuremath{\mathnormal{\alpha}}} %
\newunicodechar{β}{\ensuremath{\beta}} %
\newunicodechar{γ}{\ensuremath{\mathnormal{\gamma}}} %
\newunicodechar{δ}{\ensuremath{\mathnormal{\delta}}} %
\newunicodechar{ε}{\ensuremath{\mathnormal{\varepsilon}}} %
\newunicodechar{ζ}{\ensuremath{\mathnormal{\zeta}}} %
\newunicodechar{η}{\ensuremath{\mathnormal{\eta}}} %
\newunicodechar{θ}{\ensuremath{\mathnormal{\theta}}} %
\newunicodechar{ι}{\ensuremath{\mathnormal{\iota}}} %
\newunicodechar{κ}{\ensuremath{\mathnormal{\kappa}}} %
\newunicodechar{λ}{\ensuremath{\mathnormal{\lambda}}} %
\newunicodechar{μ}{\ensuremath{\mathnormal{\mu}}} %
\newunicodechar{ν}{\ensuremath{\mathnormal{\mu}}} %
\newunicodechar{ξ}{\ensuremath{\mathnormal{\xi}}} %
\newunicodechar{π}{\ensuremath{\mathnormal{\pi}}} %
\newunicodechar{ρ}{\ensuremath{\mathnormal{\rho}}} %
\newunicodechar{σ}{\ensuremath{\mathnormal{\sigma}}} %
\newunicodechar{τ}{\ensuremath{\mathnormal{\tau}}} %
\newunicodechar{φ}{\ensuremath{\mathnormal{\varphi}}} %
\newunicodechar{χ}{\ensuremath{\mathnormal{\chi}}} %
\newunicodechar{ψ}{\ensuremath{\mathnormal{\psi}}} %
\newunicodechar{ω}{\ensuremath{\mathnormal{\omega}}} %
\newunicodechar{ϕ}{\ensuremath{\mathnormal{\phi}}} %
\newunicodechar{ϵ}{\ensuremath{\mathnormal{\epsilon}}} %
\newunicodechar{ᵏ}{^k}
\newunicodechar{ᵢ}{_i} %
\newunicodechar{•}{\ensuremath{\cdot}} %
\newunicodechar{ }{{\,}}
\newunicodechar{′}{\ensuremath{^\prime}}  %
\newunicodechar{″}{\ensuremath{^\second}} %
\newunicodechar{‴}{\ensuremath{^\third}}  %
\newunicodechar{ⁱ}{^i}
\newunicodechar{⁵}{\ensuremath{^5}}
\newunicodechar{⁺}{\ensuremath{^+}} 
\newunicodechar{⁻}{\ensuremath{^-}} 
\newunicodechar{ⁿ}{^n}
\newunicodechar{₀}{\ensuremath{_0}} %
\newunicodechar{₁}{\ensuremath{_1}}
\newunicodechar{₂}{\ensuremath{_2}}
\newunicodechar{₃}{_3}
\newunicodechar{₊}{\ensuremath{_+}} 
\newunicodechar{₋}{\ensuremath{_-}} 
\newunicodechar{ₙ}{_n}
\newunicodechar{ₚ}{_p} %
\newunicodechar{ℂ}{\ensuremath{\mathds{C}}} %
\newunicodechar{ℕ}{\mathbb{N}} %
\newunicodechar{ℝ}{\ensuremath{\mathbb{R}}} %
\newunicodechar{ℤ}{\ensuremath{\mathbb{Z}}} %
\newunicodechar{ℳ}{\mathcal{M}}
\newunicodechar{⅋}{\ensuremath{\parr}} %
\newunicodechar{←}{\ensuremath{\leftarrow}} %
\newunicodechar{↑}{\ensuremath{\uparrow}} %
\newunicodechar{↓}{\ensuremath{\downarrow}} %
\newunicodechar{→}{\ensuremath{\rightarrow}} %
\newunicodechar{↔}{\ensuremath{\leftrightarrow}} %
\newunicodechar{↖}{\nwarrow} %
\newunicodechar{↗}{\nearrow} %
\newunicodechar{↝}{\ensuremath{\rightsquigarrow}}
\newunicodechar{↦}{\ensuremath{\mapsto}}
\newunicodechar{↭}{\ensuremath{\leftrightsquigarrow}} %
\newunicodechar{↾}{\ensuremath{\upharpoonright}} %
\newunicodechar{⇆}{\ensuremath{\leftrightarrows}} %
\newunicodechar{⇐}{\ensuremath{\Leftarrow}} %
\newunicodechar{⇒}{\ensuremath{\Rightarrow}} %
\newunicodechar{⇔}{\ensuremath{\Leftrightarrow}} %
\newunicodechar{∀}{\ensuremath{\forall}}   %
\newunicodechar{∃}{\ensuremath{\exists}} %
\newunicodechar{∅}{\ensuremath{\varnothing}} %
\newunicodechar{∈}{\ensuremath{\in}}
\newunicodechar{∉}{\ensuremath{\not\in}} %
\newunicodechar{∋}{\ensuremath{\ni}}  %
\newunicodechar{∎}{\ensuremath{\qed}}
\newunicodechar{∏}{\prod}
\newunicodechar{∑}{\sum}
\newunicodechar{∗}{\ensuremath{\ast}} %
\newunicodechar{∘}{\ensuremath{\circ}} %
\newunicodechar{∙}{\ensuremath{\bullet}} %
\newunicodechar{∙}{\ensuremath{\cdot}}
\newunicodechar{∞}{\ensuremath{\infty}} %
\newunicodechar{∣}{\ensuremath{\mid}} %
\newunicodechar{∧}{\ensuremath{\wedge}}%
\newunicodechar{∨}{\ensuremath{\vee}}%
\newunicodechar{∩}{\ensuremath{\cap}} %
\newunicodechar{∪}{\ensuremath{\cup}} %
\newunicodechar{∫}{\int}
\newunicodechar{∷}{::} %
\newunicodechar{∼}{\ensuremath{\sim}} %
\newunicodechar{≃}{\ensuremath{\simeq}} %
\newunicodechar{≅}{\ensuremath{\cong}} %
\newunicodechar{≈}{\ensuremath{\approx}} %
\newunicodechar{≔}{:=} %
\newunicodechar{≜}{\ensuremath{\stackrel{\scriptscriptstyle {\triangle}}{=}}}
\newunicodechar{≜}{\triangleq}
\newunicodechar{≝}{\ensuremath{\stackrel{\scriptscriptstyle {\text{def}}}{=}}}
\newunicodechar{≟}{\ensuremath{\stackrel {_\text{\textbf{?}}}{\text{\textbf{=}}\negthickspace\negthickspace\text{\textbf{=}}}}}
\newunicodechar{≠}{\ensuremath{\neq}}%
\newunicodechar{≡}{\ensuremath{\equiv}}%
\newunicodechar{≤}{\ensuremath{\leq}} %
\newunicodechar{≥}{\ensuremath{\geq}} %
\newunicodechar{≺}{\ensuremath{\prec}}
\newunicodechar{≼}{\ensuremath{\preceq}}
\newunicodechar{⊂}{\ensuremath{\subset}} %
\newunicodechar{⊃}{\ensuremath{\supset}} %
\newunicodechar{⊆}{\ensuremath{\subseteq}} %
\newunicodechar{⊇}{\ensuremath{\supseteq}} %
\newunicodechar{⊎}{\ensuremath{\uplus}} %
\newunicodechar{⊑}{\ensuremath{\sqsubseteq}} %
\newunicodechar{⊒}{\ensuremath{\sqsupseteq}} %
\newunicodechar{⊓}{\ensuremath{\sqcap}} %
\newunicodechar{⊔}{\ensuremath{\sqcup}} %
\newunicodechar{⊕}{\ensuremath{\oplus}} %
\newunicodechar{⊗}{\ensuremath{\otimes}} %
\newunicodechar{⊞}{\ensuremath{\boxplus}} %
\newunicodechar{⊟}{\ensuremath{\boxminus}} %
\newunicodechar{⊡}{\ensuremath{\boxdot}} %
\newunicodechar{⊢}{\ensuremath{\vdash}} %
\newunicodechar{⊣}{\ensuremath{\dashv}}
\newunicodechar{⊤}{\ensuremath{\top}}%
\newunicodechar{⊥}{\ensuremath{\bot}} 
\newunicodechar{⊧}{\vDash} %
\newunicodechar{⊨}{\vDash} %
\newunicodechar{⊩}{\Vdash}
\newunicodechar{⊸}{\ensuremath{\multimap}} %
\newunicodechar{⋁}{\ensuremath{\bigvee}}
\newunicodechar{⋃}{\ensuremath{\bigcup}} %
\newunicodechar{⋄}{\ensuremath{\diamond}} %
\newunicodechar{⋆}{\ensuremath{\star}} %
\newunicodechar{⋮}{\ensuremath{\vdots}} %
\newunicodechar{⋯}{\ensuremath{\cdots}} %
\newunicodechar{─}{---}
\newunicodechar{□}{\ensuremath{\Box}} %
\newunicodechar{▹}{\ensuremath{\rhd}} %
\newunicodechar{◂}{\ensuremath{\triangleleft}}
\newunicodechar{◃}{\triangleleft{}} %
\newunicodechar{◆}{\blacklozenge} %
\newunicodechar{◇}{\ensuremath{\Diamond}} %
\newunicodechar{◽}{\ensuremath{\square}} %
\newunicodechar{★}{\ensuremath{\star}}   %
\newunicodechar{♢}{\ensuremath{\diamondsuit}} %
\newunicodechar{♭}{\ensuremath{\flat}} %
\newunicodechar{♯}{\ensuremath{\sharp}} %
\newunicodechar{✓}{\ensuremath{\checkmark}} %
\newunicodechar{⟂}{\ensuremath{\bot}} 
\newunicodechar{⟦}{\ensuremath{\llbracket}}
\newunicodechar{⟧}{\ensuremath{\rrbracket}}
\newunicodechar{⟨}{\ensuremath{\langle}} %
\newunicodechar{⟩}{\ensuremath{\rangle}} %
\newunicodechar{⟶}{\ensuremath{\longrightarrow}}
\newunicodechar{⟷}{\longleftrightarrow}
\newunicodechar{⟹}{\ensuremath{\Longrightarrow}} %
\newunicodechar{⦃}{\ensuremath{\llbrace}} %
\newunicodechar{⦄}{\ensuremath{\rrbrace}} %
\newunicodechar{⧟}{\ensuremath{\bidirmultimap}} %
\newunicodechar{⨁}{\bigoplus} %
\newunicodechar{⨂}{\bigotimes} %
\newunicodechar{⫪}{\ensuremath{\top\kern-1.3ex\top}}
\newunicodechar{ⱼ}{_j} %
\newunicodechar{𝒟}{\ensuremath{\mathcal{D}}} %
\newunicodechar{𝒢}{\ensuremath{\mathcal{G}}}
\newunicodechar{𝒦}{\ensuremath{\mathcal{K}}} %
\newunicodechar{𝔸}{\ensuremath{\mathds{A}}} %
\newunicodechar{𝔹}{\ensuremath{\mathds{B}}} %
\newunicodechar{𝟘}{\ensuremath{\mymathds{0}}} %
\newunicodechar{𝟙}{\ensuremath{\mymathds{1}}} %

\title{A Graded Modal Type Theory for Pulse Schedules}
\author{Robin Adams}{Chalmers University of Technology, Gothenburg, Sweden \and University of Gothenburg, Sweden}{robinad@chalmers.se}{https://orcid.org/0000-0003-2644-1093}{}
\author{Jean-Philippe Bernardy}{Chalmers University of Technology, Gothenburg, Sweden \and University of Gothenburg, Sweden}{bernardy@chalmers.se}{https://orcid.org/0000-0002-8469-5617}{}
\author{Lorenzo Perticone}{Chalmers University of Technology, Gothenburg, Sweden \and University of Gothenburg, Sweden}{lorenzop@chalmers.se}{https://orcid.org/0000-0003-0198-7022}{}
\author{Jeremy Pope}{Chalmers University of Technology, Gothenburg, Sweden \and University of Gothenburg, Sweden}{popje@chalmers.se}{}{}
\authorrunning{R. Adams, J. P. Bernardy, L. Perticone, J. Pope}
\Copyright{Robin Adams, Jean-Philippe Bernardy, Lorenzo Perticone, and Jeremy Pope}

\funding{This research was funded by the SSF (Swedish Foundation for Strategic Research) project QuantumStack, grant number FUS21-0063.}

\supplementdetails{Software}{https://codeberg.org/radams78/grampus.git}

\newcommand\longversion[2]{#2}
\newcommand{\lett}{\ensuremath{\operatorname{let}}}
\newcommand{\inn}{\ensuremath{\operatorname{in}}}
\newcommand\ann[1]{\textcolor{blue}{#1}}
\newcommand\cdur[1]{^{\ann{#1}}}
\newcommand{\boxdur}[1]{\ann{\boxed{#1}}}
\newcommand\erase[1]{\lfloor #1 \rfloor}
\newcommand\Shift[1]{\mathsf{Shift}(#1)}
\newcommand{\jpb}[1]{\todo[linecolor=purple,backgroundcolor=purple!25,bordercolor=purple]{#1}}
\newcommand{\lp}[1]{\todo[linecolor=green,backgroundcolor=green!25, bordercolor=green]{#1}}

\theoremstyle{definition}
\newtheorem{df}[theorem]{Definition}

\begin{document}

\maketitle

\begin{abstract}
    The operations to be performed by a quantum computer are almost invariably given in the form of a quantum circuit. In the final stage of compilation, a quantum circuit must be translated into the input signals accepted by the quantum hardware itself. For a quantum computer based on superconducting qubits, this will be a sequence of microwave control pulses to be sent to the various input channels. A \emph{pulse schedule} gives a full specification for which pulse should be applied to which channel at what time. There is as yet no language for these pulse schedules that is very amenable to formal semantics.

In this paper, we propose such a language called GRAMPUS (GRAded Modal type theory for PUlse Schedules). It is a \emph{graded modal type theory}, where the grades represent timing information: a variable $x :^{50} Q_1$ will represent a state of qubit $Q_1$ that will exist 50 nanoseconds in the future, and a variable $y :^{-75} Q_2$ will represent a state of qubit $Q_2$ that existed 75 nanoseconds in the past.

We give the syntax for two type theories, one with grades (the \emph{annotated} language) and one without (the \emph{plain} language). We prove some metatheoretic properties, and describe the semantics in terms of category theory. We show that the input signals to a quantum chip forms a model of the annotated language. We also give a syntatic model, prove that it is initial, and hence prove soundness and completeness theorems.
\end{abstract}


    
  
  
\section{Introduction}\label{sec:intro}

The operations to be performed by a quantum computer are almost invariably given in the form of a \emph{quantum circuit} (see e.g. \cite[Ch. 4]{nielsen00}). This is a well-defined language with universally agreed semantics: there is no ambiguity about the unitaries and measurement operations that a quantum circuit diagram denotes.

In the final stage of compilation, a quantum circuit must be translated into the input signals accepted by the quantum hardware itself. For a quantum computer based on superconducting qubits, this will be a sequence of microwave control pulses to be sent to the various input channels \cite{10.1063/1.5089550}. A \emph{pulse schedule} gives a full specification for which pulse should be applied to which channel at what time.

There are languages for describing these pulse schedules, such as OpenPulse \cite{mckay2018qiskitbackendspecificationsopenqasm, Alexander_2020}. However these languages are not very amenable to formal semantics. If we want, for example, a formally verified quantum compiler, then we require a language for specifying pulse schedules that can be given precisely defined semantics.

We propose a type theory called GRAMPUS (GRAded Modal type theory for PUlse Schedules). It is a linear type theory augmented with \emph{grades}~\cite{brunel_core_2014,abel_unified_2020} (equivalently modalities) which represent relative timing information: a variable $x :^{50} Q_1$ will represent a state of qubit $Q_1$ that will exist 50 nanoseconds in the future, and a variable $y :^{-75} Q_2$ will represent a state of qubit $Q_2$ that existed 75 nanoseconds in the past.
Such a language can then be compiled to a pulse schedule. The correctness of this compiler can be captured by the following commutation square:
\vspace{-1ex}
\begin{equation}
\label{eq:main-diagram}
\begin{tikzcd}[column sep=10em,row sep=3em]
  Γ ⊢_{\ann d} t : A \arrow[d, "⟦·⟧_p"]  \arrow[r, "\erase{\cdot}"] & Γ ⊢ t : A \arrow[d, "⟦·⟧_u"] \\
 Time → Channel → \mathbb{R} \arrow[r, "\text{hardware model}"] &  ℂ^{2^n} ⊸ ℂ^{2^n}
\end{tikzcd}
\end{equation}

In the above, $\Gamma \vdash t : A$ represents a \emph{plain program} that corresponds to a quantum circuit without measurement (and with no timing information). The judgement $Γ ⊢_d t : A$ represents a quantum circuit annotated with timings. Such annotations can be erased ($\erase{\cdot}$) to obtain a plain program $Γ ⊢ t : A$.
In the semantics, a pulse schedule maps every channel and every time to a signal: $(Channel → Time → \mathbb{R})$. We denote the space of unitary operators as \(ℂ^{2^n} ⊸ ℂ^{2^n}\).

The plain program can be interpreted to a unitary operator. The hardware model maps a pulse schedule to the unitary operator that it implements.
The interpretation function $⟦·⟧_u$ maps a program $Γ ⊢ t : A$ to a
unitary operator in $⟦Γ⟧_u ⊸ ⟦A⟧_u$ (Section \ref{sec:sem-unitary}).  The
interpretation function $⟦·⟧_p$ maps an annotated program
$Γ ⊢_d t : A$ into a global pulse schedule, by combining the pulses
associated with each gate occurring in the program (Section \ref{sec:sem-pulses}).

If the hardware model is correct then the diagram commutes (Theorem \ref{thm:correctness}), ensuring that a pulse schedule implements the same unitary operator as its corresponding circuit.

\section{Syntax of the Programming Languages}
A quantum chip consists of a number of \emph{qubits} (quantum bits) with input channels (or drive lines) and output channels (or readout lines). There are certain operations or \emph{quantum gates} that we wish to perform on the qubits. We assume given a number of gates (usually called the \emph{native gates} for the chip) such that we know the signal (or pulse) that must be sent down the appropriate input channel(s) in order to effect that gate.

Our languages are based on the multiplicative fragment of (first-order) intuitionistic linear logic \cite{10.1007/3-540-56992-8_6}.
Namely, we have products $(A \otimes B)$ with unit $1$.
Additionally, we have a base type $q_i$ for each qubit $q_i$, representing the state of that qubit.
We can also annotate a type with a (relative) timing ($\boxdur d A$) meaning the data will be available exactly $d$ units of time in the future.
In the annotated language, every variable in the context will be annotated with its relative availability. In what follows, the part of the syntax related to annotations is shown in blue. The erasure function $\erase{\cdot}$ removes exactly this part.

We assume given:
\begin{itemize}
    \item a set $\mathbf{Q}$ of \emph{qubits} (or \emph{qubit labels})
    \item a set $\mathbf{G}$ of \emph{gates} (or \emph{gate constants})
    \item for every gate $G \in \mathbf{G}$, a tuple $(q_1, \ldots, q_n)$ of distinct elements of $\mathbf{Q}$ which the gate acts on, and a non-negative real number $d_G$, the \emph{duration} of the gate $G$. We say that a gate $G$ is an \emph{$n$-ary gate} iff its tuple of qubits has length $n$. We shall often write $(G, d_G, q_1, \ldots, q_n) \in \mathbf{G}$.
\end{itemize}

The syntax of GRAMPUS is then given by the grammar:
\begin{align*}
   \text{Type}\ & A,B & ::=\ &1 \mid q \mid A \otimes B \mid \boxdur{d} A \\
   \text{Context}\ & \Gamma & ::=\ & [] \mid \Gamma, x : \cdur d A \\
   \text{Term}\ & s,t & ::=\ & x \mid \star \mid \lett \star = s \inn t \mid G(t_1, \ldots, t_n) \mid (s,t) \\
                & & \mid\ & \lett\ (x,y) = s \inn t  \mid \ann{\text{box}\ d}\ t \mid \text{let }\ \ann{\text{box}\ d}\ x = s \text{ in } t  \mid \ \ann{\text{delay}_d(}t\ann{)}\\
  & & \mid & \lett x = s \inn u \\
  \text{Judgement}\ & \mathcal{J} & ::=\ & \Gamma \vdash_{\ann d} t : A \mid \Gamma \vdash_{\ann d} s = t : A
\end{align*}
where $q \in \mathbf{Q}$, $d \in \mathbb{R}$ and $G \in \mathbb{G}$ is an $n$-ary gate.

\begin{example}
Consider a very simple quantum chip with two qubits $q_1$ and $q_2$, where the native gates are:
\begin{itemize}
    \item $H_1$, a Hadamard gate applied to $q_1$ with duration 100 nanoseconds;
    \item $H_2$, a Hadamard gate applied to $q_2$ with duration 110 nanoseconds;
    \item $CNOT$, a controlled NOT gate applied to $q_1$ and $q_2$ with duration 500 nanoseconds
\end{itemize}
That is, $\mathbf{Q} = \{q_1, q_2\}$ and $\mathbf{G} = \{ (H_1, 100, q_1), (H_2, 110, q_2), (CNOT, 500, q_1, q_2) \}$.

We wish to run the following circuit on this chip:

\Qcircuit{ & \gate{H} & \ctrl{1} & \qw \\ & \gate{H} & \targ & \qw }

In the plain language, this is represented by the term
\[ x : q_1, y : q_2 \vdash CNOT(H_1(x),H_2(y)) : q_1 \otimes q_2 \]
In the annotated language, we can represent this as a pulse schedule where $q_1$'s timeline starts at time $-100$ and $q_2$'s timeline starts at time $-110$ as follows:
\[ x :^{-100} q_1, y :^{-110} q_2 \vdash CNOT(H_1(x),H_2(y)) : q_1 \otimes q_2 \]
Or we can represent the pulse schedule where we apply the Hadamard gate to both qubits starting at time $-110$, and insert a delay of 10 nanoseconds after the Hadamard gate on qubit $q_1$, as follows:
\[ x :^{-110} q_1, y :^{-110} q_2 \vdash CNOT(H_1(\text{delay}_{10}(x)), H_2(y)) : q_1 \otimes q_2 \]
In both of these examples, we start the pulse schedule at a negative time, and then end at time zero. We can shift the second example so it starts at time 0 and ends at time 110 as follows:
\[ x :^0 q_1, y :^0 q_2 \vdash \text{box}\ 110\ CNOT(H_1(\text{delay}_10(x)), H_2(y)) : \boxdur{110} (q_1 \otimes q_2) \]
\end{example}

As expected, terms of the language allow manipulating products, but we can also apply an $n$-ary quantum gate $G$ operating on an $n$-tuple of the qubits, or apply a delay of any duration to any qubit, which represents sending no input signal for that period of time. (In the ideal quantum computer, applying a delay does not change the qubit's state. In a real quantum computer, the state of an idle qubit quickly decays as it loses energy to, and gains noise from, the environment.)

The typing rules for the language are given in Figure \ref{figure:rules-of-deduction}. The key point is that if a gate $G$ takes $d$ units of time to perform its operation in hardware, then we annotate the corresponding inputs by $-d$, capturing the fact that to obtain the expected output \emph{now}, the inputs must be provided $d$ units of time in the past. 
The introduction and elimination rules for $\boxdur d A$ are setup in a way to correctly represent a time-shift{ by} $d$, following the pattern of general frameworks for grades or modalities \cite{abel_unified_2020,brunel_core_2014}.
It is easy to see that erasure preserves typing.
\providecommand\rul[1]{\text{#1}}
\begin{figure}\small
  \begin{mathpar}
    \inferrule{\Gamma, x: \cdur d A, y :\cdur e B, \Delta \vdash t : C}{\Gamma, y :\cdur e B, x :\cdur d A, \Delta \vdash t : C}\rul{Swap}
    \and
\inferrule{~}{x :\cdur 0 A \vdash x : A}\rul{Var}
    \and
\inferrule{~}{\vdash * : 1}\rul{Unit-I}
    \and
\inferrule{\Gamma \vdash s : 1\\\Delta \vdash t : A}{\ann{d +} \Gamma, \Delta \vdash \lett * = s \inn t : A}
\rul{Unit-E}
\and
\inferrule{\Gamma_1 \vdash t_1 : q_1\\ \cdots \\ \Gamma_n \vdash t_n : q_n}{\ann{-d} + \Gamma_1, \ldots, \ann{-d}+\Gamma_n \vdash G(t_1, \ldots, t_n) : q_1 \otimes \cdots \otimes q_n}\rul{Gate}
\and
\inferrule{\Gamma \vdash s : A\\\Delta \vdash t : B}{\Gamma, \Delta \vdash (s,t) : A \otimes B}\rul{Tens-I}
\and
\inferrule{\Gamma \vdash s : A \otimes B\\ x : \cdur d A, y : \cdur d B, \Delta \vdash t : C}{\ann {d +} \Gamma, \Delta \vdash \lett (x,y) = s \inn t : C} \rul{Tens-E}
\and
\inferrule{\Gamma \vdash t : A}{d + \Gamma \vdash \text{box}\ d\ t : \boxdur{d} A}\rul{Box-I}
\and
\inferrule{\Gamma \vdash s : \boxdur{d} A\\ x : \cdur e A, \Delta \vdash t : B} {\ann {e - d + }\Gamma, \Delta \vdash \text{let box} \ d\ x = s \text{ in } t : B} \rul{Box-E}
\and
    \inferrule{\Gamma \vdash t : q \\ \ann{d} ≥ 0}{\ann{- d + } \Gamma \vdash \ann{\text{delay}_d(}t\ann{)} : q} \rul{Delay}
\and
\inferrule{\Gamma \vdash s : A\\ x : \cdur d A,\Delta \vdash t : C}{\ann {d +} \Gamma, \Delta \vdash \lett x = s \inn t : C} \rul{Let}
\end{mathpar}
\caption{Typing rules for our languages.
  In the \rul{Gate} rule, $G$ is a gate acting on $(q_1, \ldots, q_n)$ with duration $\ann d$.
  Timing annotations are highlighted in blue.
We write $\ann d + \Gamma$ for the result of increasing every grade in the context $\Gamma$ by $d$.}
\label{figure:rules-of-deduction}
\end{figure}

\subsection{Judgemental Equality}

We have rules establishing that judgemental equality is reflexive, symmetric, transitive, and a congruence for each of the primitive constructors, plus:

\paragraph*{Beta Rules}

\begin{prooftree}
    \AxiomC{$\Gamma \vdash t : A$}
    \UnaryInfC{$\Gamma \vdash (\text{let } * = * \text{ in } t) = t : A$}
\end{prooftree}

\begin{prooftree}
    \AxiomC{$\Gamma \vdash s : A$}
    \AxiomC{$\Delta \vdash t : B$}
    \AxiomC{$x :^d A, y :^d B, \Theta \vdash u : C$}
    \TrinaryInfC{$d + \Gamma, d + \Delta, \Theta \vdash (\text{let } (x,y) = (s,t) \text{ in } u) = u[x:=s,y:=t] : C$}
\end{prooftree}

\begin{prooftree}
    \AxiomC{$\Gamma \vdash s : A$}
    \AxiomC{$x :^e A, \Delta \vdash t : B$}
    \BinaryInfC{$e + \Gamma, \Delta \vdash (\text{let box}\ d\ x = \text{box}\ d\ s \text{ in } t) = t[x:=s] : B$}
\end{prooftree}

\begin{prooftree}
    \AxiomC{$\Gamma \vdash s : A$}
    \AxiomC{$x :^e A, \Delta \vdash t : B$}
    \BinaryInfC{$e + \Gamma, \Delta \vdash (\text{let}\ x = s \text{ in } t) = t[x:=s] : B$}
\end{prooftree}

\paragraph*{Eta Rules}

\begin{prooftree}
    \AxiomC{$\Gamma \vdash t : I$}
    \UnaryInfC{$\Gamma \vdash (\text{let } * = t \text{ in } *) = t : I$}
\end{prooftree}

\begin{prooftree}
    \AxiomC{$\Gamma \vdash t : A \otimes B$}
    \UnaryInfC{$\Gamma \vdash (\text{let } (x,y) = t \text{ in } (x,y)) = t : A \otimes B$}
\end{prooftree}

\begin{prooftree}
    \AxiomC{$\Gamma \vdash t : \boxed{d} A$}
    \UnaryInfC{$\Gamma \vdash (\text{let box}\ d\ x = t \text{ in box}\ d\ x) = t : \boxed{d} A$}
\end{prooftree}

Note we do not have $t = *$ for $t : 1$. (If $t$ represents a process that takes time $> 0$ then it is not semantically equal to $*$.)

\paragraph*{Commuting Conversions}

In any linear type theory we also need judgemental equality rules for \emph{commuting conversions}. The full list is given in Appendix \ref{appendix:comcon}.

\section{Semantics}\label{sec:models}

In this section, we describe the interpretation for our languages.  We first introduce
algebraic structures capturing what is needed to interpret such
languages (models), and describe a general procedure to construct
interpretations which we shall employ throughout this section. In the
following, we assume fixed a set of qubit labels $\mathbf{Q}$ and a
set of gates $\mathbf{G}$, whose elements $(G, d_G, q_1, \dots, q_n)$
encode that gate $G$ acts on qubits $(q_1, \dots, q_n)$ in $d_G$ units
of time.

\subsection{Categorical model of the plain language}\label{sec:sem-categorical-plain}

The plain language can be interpreted in arbitrary symmetric monoidal
categories, using the ideas from \cite{brunel_core_2014}.

\begin{df}[Categorical model of the plain language]\label{def:PLAPUS-model}
A \emph{(categorical) model} for the plain language $(\mathcal{C}^\otimes, X_{-}, f_{-})$ is given by the following data:
\begin{itemize}
\item A symmetric monoidal category (SMC) $\mathcal{C}^\otimes$,
\item For every qubit $q \in \mathbf{Q}$, an object $X_q : \mathcal{C}^\otimes$,
\item For every gate $(G, d, q_1, \dots, q_n)$ a morphism $f_G : \bigotimes_{i = 1}^{n} X_{q_i} \to \bigotimes_{i = 1}^{n} X_{q_i}$
\end{itemize}
\end{df}

We can hence interpret the plain language in any of its categorical models: the interpretation is constructed as follows.

\begin{df}[Interpretation of the plain language]\label{def:PLAPUS-interpretation}
Given a categorical model of the plain language $(\mathcal{C}^\otimes,X_{-}, f_{-})$, we define the following interpretation $⟦-⟧_u$ operation:
\begin{itemize}
\item For every type $A$, we define an object $⟦A⟧_u \in \mathcal{C}^\otimes$; we set $⟦1⟧_u := I$, $⟦q⟧_u := X_q$ and $⟦A \otimes B⟧_u := ⟦A⟧_u \otimes ⟦B⟧_u$;
\item For every context $\Gamma$, we define an object $⟦\Gamma⟧_u$; we set $⟦[]⟧_u := I$ and $⟦x : A, \Gamma⟧_u := ⟦A⟧_u \otimes ⟦\Gamma⟧_u$;
\item For every derivable judgement $\Gamma \vdash t : A$, we define a morphism $⟦\Gamma \vdash t : A⟧_u := ⟦\Gamma⟧_u \xrightarrow{⟦t⟧_u} ⟦A⟧_u$, where $⟦t⟧_u$ is defined by induction on the judgement's derivation: Var and Unit-I correspond to identities, Swap is composition with $\mathcal{C}^\otimes$'s symmetry, Unit-E and Tens-I are the obvious tensor products, Tens-E becomes the following morphism:
\[⟦\lett (x , y) = s \inn t⟧_u = ⟦\Gamma⟧_u \otimes ⟦\Delta⟧_u \xrightarrow{⟦s⟧_u \otimes \mathbf{Id}} ⟦A⟧_u \otimes ⟦B⟧_u \otimes ⟦\Delta⟧_u \xrightarrow{⟦t⟧_u} ⟦C⟧_u\]
and Gate becomes the following morphism:
\[⟦G(t_1, \dots, t_n)⟧_u = \bigotimes_{i = 1}^{n} ⟦\Gamma_i⟧_u \xrightarrow{\bigotimes_{i = 1}^{n} ⟦t_i⟧_u} \bigotimes_{i = 1}^{n} ⟦q_i⟧_u \xrightarrow{f_G} \bigotimes_{i = 1}^{n} ⟦q_i⟧_u\]
\end{itemize}
\end{df}

\begin{proposition}\label{lem:PLAPUS-interpretation-preserves-equalities} The interpretation defined above preserves judgemental equality: if the judgment $s = t$ is admissible, then $⟦s⟧_u = ⟦t⟧_u$.
\end{proposition}

\begin{proof} This is straightforward to prove by induction on the derivation of the equality judgement.
\end{proof}

\subsection{Categorical model of the annotated language}\label{sec:sem-categorical-annotated}

In analogy to the previous subsection, we can interpret the annotated
language in (appropriately structured) SMCs, which we describe
here. Most of our definitions are analogous to what we described for
the plain language.  We additionally introduce a symmetric monoidal endofunctor
$\Shift{x}$ for any real number $x \in \mathbb{R}$, to interpret time shifts
(through their action on objects, hence on types and
contexts) and the box construct (through their action on morphisms,
hence on terms). Symmetry and monoidality for such endofunctors
corresponds to the grades interacting appropriately with tensor
products and unit types (cf. the commuting conversions,
Appendix \ref{appendix:comcon}).


\begin{df}[Categorical model of the annotated language]\label{def:GRAMPUS-model}
A categorical model of the annotated language $(\mathcal{C}^\otimes, X_{-}, f_{-}, \Shift{-}, \text{delay}_{-})$ is given by the following:
\begin{itemize}
  \item An SMC $\mathcal{C}^\otimes$,
  \item For every qubit $q \in \mathbf{Q}$ an object $X_q : \mathcal{C}^\otimes$,
  \item A monoidal functor $\Shift{-} : \mathbb{R} \to \mathbf{SMC}[\mathcal{C}^\otimes, \mathcal{C}^\otimes]$ from the discrete category of real numbers (considered as a monoidal category under $+$) to the category of symmetric monoidal endofunctors on $\mathcal{C}^\otimes$,
  \item For every gate $(G, d, q_1, \dots, q_n)$, a morphism $f_G : \Shift{-d}(\bigotimes_{i = 1}^{n} X_{q_i}) \to \bigotimes_{i = 1}^{n} X_{q_i}$,
  \item For every real number $x \in \mathbb{R}$ and qubit $q \in \mathbf{Q}$, a morphisms $\text{delay}_x : \Shift{-x}(X_q) \to X_q$,
\end{itemize}
\end{df}

The above definition is worth unpacking: monoidality for the functor $\Shift{-}$ means that, for every real numbers $d_1, d_2 \in \mathbb{R}$ and object $X : \mathcal{C}^\otimes$, we get isomorphisms (natural in $d_1, d_2, X$, and appropriately coherent)
\[ \Shift{0}(X) \simeq X\ \text{and}\ \Shift{d_1+d_2}(X) \simeq \Shift{d_1}(\Shift{d_2}(X)) \]
and for a fixed $d \in \mathbb{R}$, monoidality of $\Shift{d}$ means that, for every $X, Y : \mathcal{C}^\otimes$, we get isomorphisms (natural in $d, X, Y$ and, as before, appropriately coherent)
\[ \Shift{d}(I) \simeq I \ \text{and}\ \Shift{d}(X \otimes Y) \simeq \Shift{d}(X) \otimes \Shift{d}(Y) \]

This allows us to interpret the annotated language in any of its categorical models. The interpretation is built in a way analogous to Definition \ref{def:PLAPUS-interpretation}; the only differences relate to annotations, Box-I, Box-E and Delay (which are missing from the plain language).

\begin{df}[Interpretation for the annotated language]\label{def:GRAMPUS-interpretation}
Given a categorical model of the annotated language $(\mathcal{C}^\otimes, X_{-}, f_{-}, \Shift{-}, \text{delay}_{-})$, we define an interpretation $⟦-⟧_p$ as follows:
\begin{itemize}
\item For types we define $⟦-⟧_p$ in the same way as $⟦-⟧_u$, plus the clause $⟦\boxdur{d} A⟧_p := \Shift{d}⟦A⟧_p$,
\item For every context $\Gamma$, we define an object $⟦\Gamma⟧_p$; we set $⟦[]⟧_p := I$ and $⟦x : \cdur{d} A, \Gamma⟧_p := \Shift{d}⟦A⟧_p \otimes ⟦\Gamma⟧_p$ (notice that monoidality of \(\Shift{d}\) implies $⟦d + \Gamma⟧_p \simeq \Shift{d} ⟦\Gamma⟧_p$),
\item For judgements, we define $⟦-⟧_p$ in the same way as $⟦-⟧_u$ (after inserting coherences where appropriate); furthermore, we interpret
\[⟦\text{delay}_d(t)⟧_p = ⟦- d + \Gamma⟧_p \simeq \Shift{d}⟦\Gamma⟧_p \xrightarrow{\Shift{d} ⟦t⟧_p} \Shift{d} ⟦q⟧_p \xrightarrow{\text{delay}_x}⟦q⟧_p\]
\[⟦\text{box} \ d \ t⟧_p = ⟦d + \Gamma⟧_p \simeq \Shift{d}⟦\Gamma⟧_p \xrightarrow{\Shift{d} ⟦t⟧_p} \Shift{d} ⟦A⟧_p\]
\[⟦\text{let box} \ d \ x = s \text{ in } t⟧_p = ⟦(e - d) + \Gamma⟧_p \otimes ⟦\Delta⟧_p \xrightarrow{\phi \otimes \mathbf{Id}} \Shift{e}⟦A⟧_p \otimes ⟦\Delta⟧_p \xrightarrow{⟦t⟧_p} ⟦B⟧_p\]
where the arrow $\phi$ is defined as
\[\phi = ⟦(e - d) + \Gamma⟧_p \simeq \Shift{e - d}(⟦\Gamma⟧_p) \xrightarrow{\Shift{e - d}⟦s⟧_p} \Shift{e - d}(\Shift{d}⟦A⟧_p) \simeq \Shift{e}⟦A⟧_p\]
\end{itemize}
\end{df}
For instance, the term generated by Tens-I is interpreted as the following
\begin{align*}
    ⟦\lett (x, y) = s \inn t⟧_p & = ⟦d + \Gamma , \Delta⟧_p \\ 
    & \simeq \Shift{d}(⟦\Gamma⟧_p) \otimes ⟦\Delta⟧_p \\
    & \xrightarrow{\Shift(d)⟦s⟧_p} \Shift{d}(⟦A⟧_p \otimes ⟦B⟧_p) \otimes ⟦\Delta⟧_p \\
    & \simeq \Shift{d} ⟦A⟧_p \otimes \Shift{d} ⟦B⟧_p \otimes ⟦\Delta⟧_p \\
    & \xrightarrow{⟦t⟧_p} ⟦C⟧_p
\end{align*}
and gates become
\begin{align*}
    ⟦G(t_1, \dots, t_n)⟧_p & = ⟦- d + \Gamma_1, \dots, - d + \Gamma_n⟧_p \\
    & \simeq \Shift{d}(\bigotimes_{i = 1}^{n} ⟦\Gamma_i⟧_p) \\
    & \xrightarrow{\Shift{d}(\bigotimes_{i = 1}^{n} ⟦t_i⟧_p)} \Shift{d}(\bigotimes_{i = 1}^{n}⟦q_i⟧_p) \\
    & \xrightarrow{f_G} \bigotimes_{i = 1}^{n}⟦q_i⟧_p
\end{align*}

\begin{proposition}
\label{lem:GRAMPUS-interpretation-preserves-equalities}
The interpretation defined above preserves judgemental equality: if the judgment $s = t$ is admissible, then $⟦s⟧_u = ⟦t⟧_u$.
\end{proposition}

\begin{proof} Again, this is straightforward to prove by induction on the derivation of the equality judgement.
\end{proof}

Every model of the plain language can be made into a model of the annotated language; this corresponds to erasing timing annotations (the $\erase{-}$ operation defined in Section \ref{sec:intro}).

\begin{remark}\label{rmk:erasing-on-models}
Given a model of the plain language $(\mathcal{C}^\otimes, X_{-}, f_{-})$, we can define a model of the annotated language by defining
\begin{itemize}
\item Time shifts to be identity functors $\Shift{d} := \mathbf{Id}_{\mathcal{C}^\otimes}$ for every real number $d$, and
\item Delays to be identity morphisms $\text{delay}_d := \mathbf{Id}_{X_q}$.
\end{itemize}
\end{remark}

\subsection{Hilbert Spaces and Unitary Operators}\label{sec:sem-unitary}

Importantly, the category of finite-dimensional Hilbert spaces and unitary operators $\mathbf{FdHilb}_{\mathbf{U}}$ is a categorical model of the plain language (Definition \ref{def:PLAPUS-model}):

\begin{proposition}[Categorical model of Unitary Operators]\label{prop:PLAPUS-unitary-model}\ 
Assume that, for every $n$-ary gate $G$, we have a unitary operator $f_G : \mathbb{C}^{2^n} \rightarrow \mathbb{C}^{2^n}$.
The category $\mathbf{FdHilb}_{\mathbf{U}}$ is symmetric monoidal (i.e. $\mathbf{FdHilb}$ is symmetric monoidal and its identities, associators, unitors and symmetries are unitary operators). This, together with the assignments
\begin{itemize}
\item $X_q := ℂ^2$ for every qubit $q \in \mathbf{Q}$;
\item $f_G$ as the linear operator corresponding to the quantum gate $G$.
\end{itemize}
makes it into a categorical model of the plain language.
\end{proposition}

In consequence, programs in the plain language can be interpreted in
the ideal quantum computer --- ignoring all aspects related to
duration of execution. A program \(Γ ⊢ t : A\) hence corresponds to a unitary operator
\(⟦t⟧ ∈ ⟦Γ⟧ ⊸ ⟦A⟧\).  In particular, for any gate \(G\) operating on
\(n\) qubits, we assume given a corresponding operator
\(⟦G⟧ ∈ ℂ^{2^n} ⊸ ℂ^{2^n}\).

\subsection{Signals}\label{sec:sem-pulses}

In a physical quantum chip with superconducting qubits, there are a number of input channels or \emph{drive lines} that can be used to perform operations on the qubits, and \emph{readout lines} for receiving the outcome of measuring a qubit. (For details, see \cite{10.1063/1.5089550}.)

%

The input to a drive line can be modelled as a function $[startTime, endTime] \rightarrow \mathbb{R}$.
A complete pulse schedule consists then of such a function for each channel connected to such a drive line: $Channel \to [startTime, endTime] \to \mathbb R$.
However, these do not form the morphisms of an SMC; we can only form the tensor product of two pulse schedules if their start times and end times agree. To support the SMC structure, we must allow a different start or end time for each channel: $(c : Channel) \to [StartTime(c), EndTime(c)] \to \mathbb R$.

We can then define an SMC $\mathcal S$ whose morphisms are the pulse schedules themselves.

\begin{df}[SMC of Signals]\label{def:signals-SMC}
Given a set of channels $\mathbf{C}$, we define the SMC $\mathcal{S}$ such that:
\begin{itemize}
\item Objects are finite sequences of pairs $\mathbf{Ob}(\mathcal{S}) := (\mathbb{R} \times \mathbf{C})^{*}$ of a real number (the starting time of the signal) and a channel;
\item Given $X, Y : \mathcal{S}$ with $X = [(s_1, c_1), \dots, (s_m, c_m)]$ and $Y = [(t_1, d_1), \dots, (t_n, d_n)]$, define the homset $\mathbf{Hom}_{\mathcal{S}}[X, Y]$ as
\[\{ (\sigma,[\phi_1,\dots,\phi_n]) \ |\ \sigma \in S_n\ \text{such that}\ \forall k,\ c_k = d_{\sigma(k)} \text{ and } s_k \leq t_{\sigma(k)} \text{ and } \phi_k : [s_k, t_{\sigma(k)}) \to \mathbb{R} \}\]
if $m = n$ and as the empty set otherwise, where $S_n$ is the symmetric group. One could require some additional regularity conditions on the functions $\phi_k$, which we simply omit;
\item Identities are hence given by pairs $(e, [\emptyset, \dots, \emptyset])$ where $e$ is the identity permutation, and $\emptyset$ is the empty function;
\item Composition is given by composing the permutations and concatenating the pulses:
\[(\sigma_1, [\phi_1, \dots, \phi_n]) \circ (\sigma_2, [\psi_1, \dots, \psi_n]) := (\sigma_1 \circ \sigma_2, [\gamma_1, \dots, \gamma_n])\]
where $\gamma_k : [r_k, t_{(\sigma_1 \circ \sigma_2)(k)}) \to \mathbb{R}$ is defined as
\begin{align*}
\gamma_k(t) := \begin{cases}
\psi_k(t) &\text{ if } t \in [r_k, s_{\sigma_2(k)}) \\
\phi_{\sigma_2(k)}(t) &\text{ if } t \in [s_{\sigma_2(k)}, t_{(\sigma_1 \circ \sigma_2)(k)})
\end{cases}
\end{align*}
(intuitively, we keep track of how channels are permuted and apply the pulses sequentially)
\item The tensor unit is given by the empty sequence $I := [] : \mathcal{S}$;
\item Tensor product of objects is given by list concatenation;
\item Tensor product of morphisms is given by the following
\[(\sigma_1, [\phi_1, \dots, \phi_m]) \otimes (\sigma_2, [\psi_1, \dots, \psi_n]) := (\sigma_1 \star \sigma_2, [\phi_1, \dots, \phi_m, \psi_1, \dots, \phi_n])\]
where by $(- \star -) : S_m \times S_n \to S_{m + n}$ we denote the obvious group homomorphism.
\item The unitors and associator are given by identities, and
\item The symmetry is given by $(\sigma, [\emptyset, \dots, \emptyset])$ where $\sigma$ is the appropriate permutation and $\emptyset$ is the empty map.
\end{itemize}
\end{df}


We can use this category to build a model of the annotated language out of the input signals to the quantum chip, in such a way that the interpretation function $⟦\ ⟧_p$ is exactly the compiler that translates from a GRAMPUS term to the corresponding input signals:

\begin{proposition}[$\mathcal{S}$ is a categorical model of the annotated language]\label{lm:GRAMPUS-signals-model}\ 

Assume given a correspondence between channels and qubits (in the simplest case this could be an injection $\mathbf{Q} \to \mathbf{C}$) and a pulse representation of each gate in $\mathbf{G}$. Then the SMC $\mathcal{S}$ can be extended to a categorical model of the annotated language as follows:
\begin{itemize}
\item For every real number $x \in \mathbb{R}$, the functor $\Shift{x}$ is defined as follows:
\[\Shift{x}[(t_1, c_1), \dots, (t_n, c_n)] := [(t_1 + x, c_1), \dots, (t_n + x, c_n)]\]
\[\Shift{x}(\sigma, [\phi_1, \dots, \phi_n]) := (\sigma, [\phi_1(\cdot - x), \dots, \phi_n(\cdot - x)])\]
\item For every real number $x \in \mathbb{R}$, the morphisms $\text{delay}_x$ are built from the identity permutation $e \in S_1$ and the pulse $\delta_0$ which is constantly zero\footnote{Notice how these morphisms are not invertible: in general, due to phenomena such as decoherence, a qubit left alone will not stay the same. For a more detailed treatment of this, see section 3 of \cite{10.1063/1.5089550}.}.
\[\text{delay}_x = (e, [\delta_0])\]
\end{itemize}
\begin{proof}
Functoriality and (symmetric) monoidality for $\Shift{x}$ are straightforward to verify. Furthermore, in this model, the structural natural isomorphisms making $\Shift{-}$ into a monoidal functor are identities: the coherence conditions listed in Definition \ref{def:GRAMPUS-model} hold trivially as a consequence.
\end{proof}
\end{proposition}

\subsection{Syntactic Model and Completeness Theorem}\label{sec:syntactic}

We can build a model for either of our type theories out of the terms of the theory itself. It is generally considered desirable if this term model is initial in the category of models, with the interpretation function $⟦\ ⟧$ being the unique morphism from the term model to an arbitrary model (see e.g. \cite{Uemura_2023}). In this section, we prove that this is the case for both of our languages.

\begin{df}[Term category for the annotated Language]\label{def:GRAMPUS-Syntax-cat}\ 

We define the model of the plain language $\mathbf{Syntax} = (\mathbf{Syntax}, X_{-}, f_{-})$ as follows. We begin by defining a function $\mathbf{Ty}$ that sends contexts $\Gamma$ to types $\mathbf{Ty}(\Gamma)$ by
\[ \mathbf{Ty}([]) := 1; \ \ \mathbf{Ty}(x : \cdur{d} A , \Gamma) := (\boxdur{d} A) \otimes \mathbf{Ty}(\Gamma) \]
We can then define the underlying category of $\mathbf{Syntax}$ as follows:
\begin{itemize}
\item Its objects are contexts,
\item Its morphisms $f : \Gamma \to \Delta$ are derivable judgments $\Gamma \vdash t : \mathbf{Ty}(\Delta)$, quotiented by judgmental equality\footnote{In the rest of this definition, we will silently treat representatives of such equivalence classes as the classes themselves.}.
\item Identities $\mathbf{id}_\Gamma : \Gamma \to \Gamma$ are defined by induction on $\Gamma$:
\[ \mathbf{Id}_{[]} := *; \ \ \mathbf{id}_{x : \cdur{d} A, \Gamma} := (x, \mathbf{id}_\Gamma) \]
\item Composition is given by inductively using the Tens-E and Unit-E rules: given $f : \Gamma \to \Delta$ and $g : \Delta \to \Xi$, if $\Delta = x_1 : \cdur{d_1} A_1 , \dots , x_n : \cdur{d_n} A_n$,
\begin{equation*}
\begin{alignedat}{2}
g \circ f :=
& \lett \ (y_1, z_1) = f \ &&\inn \ \lett \ \text{box} \ d_1 \ x_1 = y_1 \ \inn \\
& \lett \ (y_2, z_2) = z_1 \ &&\inn \ \lett \ \text{box} \ d_2 \ x_2 = y_2 \ \inn \\
& \dots \\
& \lett \ (y_n, z_n) = z_{n-1} \ &&\inn \ \lett \ \text{box} \ d_n \ x_n = y_n \ \inn \\
& \lett \  * = z_n \ &&\inn \ g
\end{alignedat}
\end{equation*}
for distinct, fresh $y_1, \dots, y_n, z_1, \dots, z_n$. We can give a derivation for the judgment $\Gamma \vdash (g \circ f) : \mathbf{Ty}(\Xi)$ by generalizing the construction:
\begin{prooftree}
    \AxiomC{$\Gamma \vdash t : \mathbf{Ty}(\Delta)$}
    \AxiomC{$\Delta, \Xi \vdash s : C$}
    \BinaryInfC{$\Gamma, \Xi \vdash \text{let}_\Delta \ t \ \inn \ s : C$}
\end{prooftree}
\end{itemize}
and performing induction on $\Delta$; the $\lett_\Delta$ statement above is defined in exactly the same way as composition, and generalizes it.
\end{df}

Verifying that the above definition indeed gives a category structure is routine: unitality and associativity of composition hinges on the usual substitution lemmas. We can further enhance the structure on $\mathbf{Syntax}$ to that of an SMC in the obvious way.

\begin{df}[Monoidal structure on $\mathbf{Syntax}$]\label{def:GRAMPUS-Syntax-SMC}\ 
The category $\mathbf{Syntax}$ can be made into an SMC $(\mathbf{Syntax}, I, \otimes, \dots)$ by defining the following:
\begin{itemize}
\item Its unit object $I$ is given by the unit type $I := 1$;
\item Its tensor product of objects is given by context concatenation $\Gamma \otimes \Delta := \Gamma , \Delta$;
\item Its tensor product of morphisms is given by inductively applying the Tens-E and Tens-I rules: if $\Gamma_1 \vdash t_1 : \mathbf{Ty}(\Delta_1)$ and $\Gamma_2 \vdash t_2 : \mathbf{Ty}(\Delta_2)$, define
\begin{equation*}
\begin{alignedat}{2}
t_1 \otimes t_2 := & \lett \ (x_1, y_1) = t_1 \ & \inn \\
& \lett \ (x_2, y_2) = y_1 \ & \inn \\
& \dots \\
& \lett \ (x_n, y_n) = y_{n-1} \ & \inn \\
& \lett \ * = y_n \ & \inn \\
& (x_1, \dots (x_n, t_2) \dots)
\end{alignedat}
\end{equation*}
\item Checking that unitors and associators can just be identities is straightforward;
\item The symmetry can be defined with a similar inductive strategy as the tensor product of morphisms (employing the Swap rule appropriately).
\end{itemize}
\end{df}

Checking the axioms of an SMC is straightforward. Since we shall evidently interpret qubits and gates as themselves (i.e. a qubit $q \in \mathbf{Q}$ is interpreted as the context $x :^0 q$, and similarly for gates), we only need to define the $\Shift{d}$ endofunctors and the $\text{delay}_d$ natural transformations.

\begin{df}[$\Shift{d}$ and $\text{delay}_d$ for the SMC $\mathbf{Syntax}$]\label{def:GRAMPUS-Syntax-modalities}\ 
For any given real number $d \in \mathbb{R}$, define the symmetric monoidal endofunctor $\Shift{d} : \mathbf{Syntax} \to \mathbf{Syntax}$ as follows:
\begin{itemize}
\item On objects $\Gamma$, define $\Shift{d}(\Gamma) := d + \Gamma$
\item On morphisms $\Gamma \vdash t : \mathbf{Ty}(\Delta)$, define
\begin{equation*}
\begin{alignedat}{2}
\Shift{d}(t) := & \lett \ (x_1, y_1) = t \ & \inn \\
& \lett \ (x_2, y_2) = y_1 \ & \inn \\
& \dots \\
& \lett \ (x_n, y_n) = y_{n-1} \ & \inn \\
& (\text{box} \ d \ x_1 , \dots (\text{box} \ d \ x_n , \text{box} \ d \ y_n) \dots )
\end{alignedat}
\end{equation*}
\end{itemize}
Endowing $\Shift{d}$ with the further structure of a symmetric monoidal functor is a simple exercise.
Similarly, given a qubit $q \in \mathbf{Q}$ and a real number $d \in \mathbb{R}$, we define $\text{delay}_d : [x : \cdur{-d} q] \to [x : \cdur{0} q]$ as (the obvious derivation for) the judgment
\[[x : \cdur{-d} q] \vdash(\text{box}\ 0\ \text{delay}_d(x), *) : q \otimes 1\]
\end{df}

The same constructions as above (Definitions \ref{def:GRAMPUS-Syntax-cat} and \ref{def:GRAMPUS-Syntax-SMC}) can be repeated for the plain language, producing an SMC $\mathbf{Syntax}_u$. By construction, $\mathbf{Syntax}$ (resp. $\mathbf{Syntax}_u$) is a model of the annotated (resp. plain) language.

\begin{remark}
Every categorical model of the plain (resp. annotated) language $(\mathcal{C}^\otimes, X_{-}, f_{-})$ (resp. $(\mathcal{C}^\otimes, X_{-}, f_{-}, \Shift{-}, \text{delay}_{-})$) admits a symmetric monoidal functor $\mathbf{Syntax}_u \to \mathcal{C}^\otimes$ preseving qubits and gates (resp. $\mathbf{Syntax} \to \mathcal{C}^\otimes$, also appropriately preserving $\Shift{-}$ and $\text{delay}_{-}$).
This functor is unique up to natural isomorphism that preserves qubits.
\end{remark}

\begin{proof}
Such a functor can be constructed by adapting Definition \ref{def:PLAPUS-interpretation} (resp. Definition \ref{def:GRAMPUS-interpretation}).
\end{proof}

\longversion{
\begin{df}
  \jpb{This is seemingly a reboot. Needs cleanup.}
  \lp{Yes, I'm keeping it here (for now) just for reference. Everything from here until the end of section 3 will be removed later on!}
The objects of $\mathbf{Syntax}$ are the types of the plain language.

Given two types $A$ and $B$, consider the set of all pairs $(x,t)$ such that
\[ Func(A,B) = \{ (x,t) \mid x : A \vdash t : B \} \enspace . \]
We define the equivalence relation $\sim$ on this set to be the equivalence relation generated by:
\begin{description}
    \item[$\alpha$-conversion] i.e. $(x,t) \sim (y, t[x:=y])$; and
    \item[judgemental equality]; if $x : A \vdash s = t : B$ then $(x,s) \sim (x,t)$
\end{description}
The set $\mathbf{Syntax}[A,B]$ of morphisms from $A$ to $B$ is the quotient set
\[ \mathbf{Syntax}[A,B] = Func(A,B) / \sim \enspace.  \]

We shall often write an element of $\mathbf{Syntax}[A,B]$ as $x : A \vdash t : B$.

The identity on $A$ is $x : A \vdash x : A$.

Composition is given by subsitution:
\[ (y : B \vdash t : C) \circ (x : A \vdash s : B) = (x : A \vdash t[y:=s] : C) \enspace . \]

The tensor product of types $A$ and $B$ is $A \otimes B$.

The tensor product on morphisms is given by:
\[ (x : A \vdash s : C) \otimes (y : B \vdash t :D) = (z : A \otimes B \vdash \text{let } (x,y) = z \text{ in } s \otimes t : C \otimes D)\]

The tensor unit is $I$.
\end{df}

\todo{
The category where
- objects are contexts
- morphisms from $Γ$ to $Δ$ is a derivation of $Γ ⊢ ⨂Δ$
form a model of the plain language.

(Define $⨂$ by induction on context)

Proof:
- existence of all morphisms (maybe the functorial action can be discussed some more).
- laws are coherent (all are easy, except maybe associativity of composition, which relies on a substitution lemma.)

Lemma: $u[x := t[y := v]]  =  u[x := t][y := v]$
Proof: by induction.}
}{}

\begin{theorem}[Completeness]
Suppose $\Gamma \vdash_{\ann d} s : A$ and $\Gamma \vdash_{\ann d} t : A$. If $⟦s⟧_p = ⟦t⟧_p)$ in every model, then $\Gamma \vdash_{\ann d} s = t : A$.
\end{theorem}

\begin{proof}
    If the equality holds in every model then it holds in $\mathbf{Syntax}$, which means precisely that $\Gamma \vdash_{\ann d} s = t : A$.
\end{proof}

A similar Completeness Theorem can be proved for the plain language.
\providecommand\Circ{\ensuremath{\mathsf{Circ}}}
\subsection{The Category $\Circ$}
\label{section:Circ}

We can give an alternative characterisation of the initial model $\mathbf{Syntax}$ as follows:
\providecommand\circuit{⊸_G}
\begin{df}\label{def:circ}
  Let $\Circ$ be the free SMC, generated as
  follows:
  \begin{itemize}
  \item Its objects are generated by qubits shifted by an arbitrary
    time. This free time shift corresponds to attaching a real
    number to each qubit, hence objects are exactly contexts whose
    types are restricted to just qubits;
  \item Its morphisms are generated by gates and delays, themselves
    possibly shifted by an arbitrary time.
  \end{itemize}
  Let us call this category circuits (\(\Circ\)) and write its homsets
  \(Γ \circuit Δ\).
\end{df}

There is a symmetric monoidal equivalence between the categories $\mathbf{Syntax}$ and $\Circ$. We can also give a similar characterisation of (a category symmetric monoidal equivalent to) the category $\mathbf{Syntax}_u$. See Appendix \ref{appendix:synmod} for more detail on both of these categories.

\section{Normal forms}

In this section we describe a \emph{normal} representation for
GRAMPUS. At the end of the section, we will deal with the
representation for $\Circ$ morphisms themselves. But we first discuss
how such morphisms should be embedded into the terms of a programming
language. We construct such normal forms in three stages:
\newcommand\elimform{⊢_e}
\newcommand\normform{⊢_n}
\begin{enumerate}
\item Decomposition of the inputs using eliminators (\(Γ \elimform C\));
\item Application of circuit gates and delays (\(Γ \circuit Δ\), see Section \ref{section:Circ});
\item Construction of the output of the desired type by using constructors
  (\(Γ \normform C\)).
\end{enumerate}
These judgements are defined by the following rules. 

\begin{mathpar}
\inferrule{Θ,Δ \elimform t : A}{Θ,x : \cdur d 1, Δ \vdash \lett * = x \inn t : A} \rul{Unit-E}
\and
\inferrule{~}{\normform * : 1}\rul{Unit-I}
\and
\inferrule{Θ,x : \cdur d A, y : \cdur d B, Δ \elimform t : C}{Θ,z : \cdur d (A⊗B), Δ \elimform \lett (x,y) = z \inn t : C} \rul{Tens-E}
\and
\inferrule{Γ \normform s : A\\Δ \normform t : B}{Γ, Δ \normform (s,t) : A ⊗ B}\rul{Tens-I}
\and
\inferrule{Θ,x : \cdur {d+e} A, Δ \elimform t : B} {Θ,z : \cdur d \boxdur e A, Δ \elimform \text{let box} \ d\ x = z \text{ in } t : B} \rul{Box-E}
\and
\inferrule{Γ \normform t : A}{d + Γ \normform \text{box}\ d\ t : \boxdur{d} A}\rul{Box-I}
\and
\inferrule{Δ \normform t : B \\ c : Θ ⊸_G Δ } {Θ \elimform t \mathsf{~after~} c : B} \rul{Circ}
\and
\inferrule{~}{x :\cdur 0 q \normform x : q}\rul{Var}
    \and
\end{mathpar}
In the above, the eliminators only work on the first non-qubit
variable in the context (\(Θ\) stands for a context with only qubits).

\begin{remark}
  Because gates have the same domain and codomain up to permutation
  and later durations, so does any circuit applied in the \rul{Circ}
  rule, and thus variable names can remain the same before and
  applying the circuit (even if they can be permuted).
\end{remark}

With the exception of the \rul{Circ} rule, the rules are a subset of
GRAMPUS, and thus are given the same semantics. With this in place, we
can show that the effect that the semantics of normal-form
introduction and elimination rules is trivial.

\begin{lemma}\label{lem:constructors-id}
If \(Γ \normform t : A\), then \(⟦Γ⟧ = ⟦A⟧\).
\end{lemma}
\begin{proof}
  The proof is a simple induction on the structure of \(t\), relying
  on the fact that the interpretation of constructors do not use gates
  or delays. Additionally \(id ⊗ id = id\) and \(\Shift{d}\ id = id\).
\end{proof}

\providecommand\normelim{⊸_e}
To state a similar result about eliminators, we want to generalise
over them. We can do so by considering the relation \(Γ \normelim Δ\), which
is any of the context transformation performed by an eliminator in the form allowed by the \(\elimform\) judgement.
\begin{mathpar}
  
  \inferrule{}{* = x ~:~ (x: \cdur d 1 \normelim [])}
  
  \inferrule{}{(x,y)=z ~:~ (z : \cdur d (A⊗B) \normelim x : \cdur d A , y : \cdur d B)}
  
  \inferrule{}{\text{box} \ d\ x = z ~:~ (z : \cdur d \boxdur e A \normelim x : \cdur {d+e} A)}
\end{mathpar}
Then all normalised eliminators as follows become an instance of the following rule
\[\inferrule{e : Γ \normelim Δ \\ Θ,Δ,Ξ ⊢ t : A}{Θ,Γ,Ξ ⊢ \lett e \inn t : A}\]
and we can state the desired lemma as follows.
\begin{lemma}\label{lem:eliminators-id}
  For any \(e : Γ \normelim Δ\), we have
  \(⟦\lett e \inn t : A⟧ = ⟦t⟧\).
\end{lemma}
\begin{proof}
  By case analysis. The fact that \(\Shift{d}\) is a monoidal functor is critical.
\end{proof}

\begin{theorem}[Normalization]Given a term \(Γ ⊢ t : A\), there is a unique $u$ such that \(Γ \elimform u: A\) and \(⟦t⟧ = ⟦u⟧\).
\end{theorem}
\begin{proof}
\begin{itemize}
\item  The circuit to embed is given by the semantics into $\Circ$;  \(⟦t⟧_{\Circ} : ⟦Γ⟧_{\Circ} \circuit ⟦A⟧_{\Circ}\) .
  (Note that in \(\Circ\), contexts evaluate to contexts.)
\item The matching constructor \(⟦A⟧_{\Circ} \normform s : A\) is
  constructed by induction on the type.  Furthermore constructor rules
  are purely type-directed, and exactly one rule applies at any point.

\item The elimination sequence which match \(Γ\) to \(⟦Γ⟧_{\Circ}\) is
  done by induction on the suffix of \(Γ\) which starts with a
  non-elementary qubit type.  As for constructors, there is a single
  rule which can apply at any point, because the first non-elementary
  type dictates which rule applies. When there are only qubits in the
  context, rule \rul{Circ} applies on \(⟦t⟧_{\Circ}\), and this is
  always applicable because the eliminators previously applied did not
  change the context semantics (Lemma \ref{lem:eliminators-id}).  Hence
  there is always exactly one eliminator to use.
\end{itemize}
The fact that this normal form respects semantics is a consequence of
Lemmata \ref{lem:constructors-id} and \ref{lem:eliminators-id}.
\end{proof}

\paragraph*{Dicussion}
We have so far established the existence of unique normal forms, but
the computation itself is grouped in single morphism of the SMC \Circ.
Finally we can consider what method should be used to best represent
morphisms in this category. There are many possible choices, because
SMCs feature a number of laws relating objects and morphisms.

One route is to map them onto the GRAMPUS syntax, such that for any
morphism \(h : Γ \circuit Δ\), give a term \(Δ,Ξ ⊢ t : C\) we can
construct a term of type \(C\) in context \(Γ,Ξ\).  The structural
morphisms correspond to context shuffling. The gates and delays must
be embedded in the following way. Let us consider the case of a gate
shifted by time \(e\), ie. mapping context \(e+Γ\) to \(e+Δ\). The corresponding term is:
\providecommand\boxx[1]{\text{box}\,#1\,}
\[\lett \boxx{e} z = \boxx{e} G(Γ) \inn \lett_{e+Δ} = z \inn t.\]
To represent composition, one can always apply one such transformation
and then the other.

Another possible choice to represent \(\Circ\) morphisms is Einstein
notation. This notation is commonly used to represent tensor
expression in physics. It is clear that plain tensors are
representable this way, but pulse schedules can be represented using
Einstein notation just as well---because every free SMC can be
represented this way. In Einstein notation, every input \emph{and}
output variable is made explicit. Every morphism is annotated with its
inputs as subscripts and outputs as superscript. So a circuit composed
of a single gate
\(H : x:q₁, y:q₂ \circuit z : \cdur d q₁, w :\cdur d q₂\) is written
\(H_{xy}^{zw}\).  The composition \(f ∘ g\) is written as a
multiplication
\(F_{inputs}^{intermediate} G_{intermediate}^{outputs}\).  Unitors,
associators and exchange morphisms are represented by the appropriate
Kronecker deltas.  Then the laws of the SMCs correspond
simplifications of such Kronecker deltas. After such simplifications,
testing the equality of morphisms in Einstein notation is a simple
syntactic test, up to renaming of intermediate variable names.  We
direct the reader to \cite{bernardy_domain-specific_2025} for details
on conversions back and forth between Einstein notation and morphisms
in an arbitrary SMC.

Yet another representation is used for scheduling (Section \ref{sec:scheduling}).

\section{Hardware Model and Correctness}
\label{sec:hwmodelcor}

A \emph{hardware model} specifies the action of a complete pulse schedule (given as a function $Time \rightarrow Channel \rightarrow \mathbb{R}$) on the state of a quantum chip. If the chip has $n$ qubits, then the state of the chip is given by a unit vector in $\mathbb{C}^{2^n}$, and inputting a pulse schedule will perform a certain unitary operator $U : \mathbb{C}^{2^n} ⊸ \mathbb{C}^{2^n}$ on the state.

In principle, we can calculate this unitary from the Hamiltonian of the chip, which is determined by the chip's circuit design. If there are $k$ input channels $V_1(t)$, \ldots, $V_k(t)$ (where $V_i(t)$ is the voltage applied on input channel $i$ at time $t$), then the Hamiltonian can be given in the form of a linear operator $H(V_1(t), \ldots, V_k(t)) : \mathbb{C}^{2^n} \rightarrow \mathbb{C}^{2^n}$.

The evolution of the state of the quantum chip is then given by $|\phi(t) \rangle = U(t) |\phi(0) \rangle$, where $|\phi(t)\rangle \in \mathbb{C}^{2^n}$ is the state at time $t$ and $U(t)$ is the unitary
\[ U(t) = \exp \left( - \frac{i}{\hbar} \int_0^t H(V_1(t'), \ldots, V_k(t')) dt' \right) \enspace . \]

Given a quantum chip whose Hamiltonian is given by the above, we say that the pulse schedule given by $V_1(t')$, \ldots, $V_k(t')$, with start time 0 and end time $t$, \emph{implements} the unitary $U(t)$.

\begin{theorem}[Correctness]
\label{thm:correctness}
Assume that we have a model of the plain language in Hilbert spaces in which every gate $G$ is assigned the unitary $U_G$; and
a model of the annotated language in $\mathcal{S}$ such that, for every gate $G$, the morphism $f_G$ is a pulse schedule that implements $U_G$.
Then the diagram (\ref{eq:main-diagram}) commutes.
\end{theorem}

\begin{proof}
The proof is by induction on the derivation of $\Gamma \vdash_d t : A$ and relies on the following observations: If pulse schedule $p$ implements $U$ and $q$ implements $V$ then $q \circ p$ implements $V \circ U$.
If $p$ implements $U$ and $q$ implements $V$ then $p \otimes q$ implements $U \otimes V$. We always have $p$ and $\mathrm{Shift}(d)(p)$ implement the same unitary.
\end{proof}

Therefore, if a scheduling algorithm (a function from the plain language to the annotated language) is a right inverse to erasure, then the pulse schedule generated implements the same unitary as the original circuit.

\section{Intermediate Representations}

\lstdefinelanguage{agda}{
    keywords=[1]{data, where}
}

\lstset{
    basicstyle=\ttfamily\footnotesize,
    extendedchars=\true,
    language=agda,
    keywordstyle=\bfseries,
    literate=
        {Σ}{{\(\Sigma\)}}1
        {λ}{{\(\lambda\)}}1
        {₂}{{\(_2\)}}1
        {≤}{{\(\le\)}}1
        {≈}{{\(\approx\)}}1
        {→}{{\(\rightarrow\)}}1
}


Intermediate representations in the dependently-typed programming language Agda~\cite{norell2007towards} are introduced in order to facilitate the implementation of scheduling; they are isomorphic respectively to the plain and annotated surface languages (\(Γ ⊢ t : A\) and \(Γ ⊢_d t : A\)) up to normalisation, but avoid redexes while being more normalised than a free SMC generated by qubits and gates.
The source code is available online;\footnote{Repository: \url{https://codeberg.org/radams78/grampus.git}} listings in this section make minor cosmetic changes to improve readability.

\subsection{Plain circuits}

The plain language is represented through the type family \lstinline{Circuit}, indexed by two lists of qubits indicating the types of the circuit's input and output.

\begin{lstlisting}[caption={Circuit},label=list,float=h]
data Circuit : Qubits → Qubits → Set where
  wire : Circuit a a
  gate : Gate a → Circuit a a
  serial : Circuit a b → Circuit b c → Circuit a c
  parallel :
    par-qubits a1 a2 a3 →
    par-qubits b1 b2 b3 →
    Circuit a1 b1 →
    Circuit a2 b2 →
    Circuit a3 b3
\end{lstlisting}

The type \lstinline{par-qubits x y z} used in the \lstinline{parallel} constructor maps each qubit to a bijection between its occurrences in \lstinline{x} or \lstinline{y} and its occurrences in \lstinline{z}.

It is defined as follows.

\begin{lstlisting}
par-qubits : Qubits → Qubits → Qubits → Set
par-qubits x y z = (q : Qubit) → Bij ((q ∈ x) ⊎ (q ∈ y)) (q ∈ z)
\end{lstlisting}
%
%
This is equivalent to permutations from the concatenation of \(x\) and \(y\) to \(z\), meaning that the \lstinline{parallel} constructor captures association, unitors, and exchange.

As is the case for the plain language, the \lstinline{Circuit} datatype does not preclude multiple occurrences of the same qubit, but is well-behaved in that it treats occurrences as distinct individuals (informally, they are never confused with one another, copied, or destroyed; the definition of \lstinline{par-qubits} plays a significant role in this).
As a result, issues of duplication can be safely confined to the observation that an element of type \lstinline{Circuit a b} is physically realizable exactly when \lstinline{a} (or equivalently, \lstinline{b}) is free of duplicates.

\subsection{Annotated circuits}


In order to describe the annotated language, the notion of a \emph{timing function} is introduced.
A timing function on a list \(x\) of qubits is a function that takes a qubit \(q\) and an element of \(q \in x\), i.e. the index of an occurrence of \(q\) in \(x\), and produces a time value:
\begin{lstlisting}[caption={Timing}, float=h]
Timing : Qubits → Set
Timing qs = (q : Qubit) → q ∈ qs → Time
\end{lstlisting}
The annotated language can then be represented in a similar way to the plain language, but with as additional type indices two timing functions \(f\) and \(g\); \(f\) sends each input qubit to the time the annotated circuit needs it, and $g$ sends each output qubit to the time the annotated circuit is done with it.

If an annotated circuit were depicted graphically as a schedule of gates, with qubits on the y-axis and time on the x-axis, the two timing functions would represent the schedule's left and right boundaries.
Accordingly, they are referred to respectively as the annotated circuit's \emph{left-timing} and \emph{right-timing}.
With respect to a typing judgement \(Γ ⊢_d t : A\), \(f\) and \(g\) capture annotations in \(Γ\) as well as boxes in \(Γ\) and \(A\).

The datatype---a possible representation of the category \Circ---is then defined as follows:

\begin{lstlisting}[caption={AC},float=h]
data AC : (a b : Qubits) → Timing a → Timing b → Set where
  delay : f ≤ g → AC a a f g
  gate :
    (gt : Gate a) →
    constantly t f →
    constantly (t + duration gt) g →
    AC a a f g
  serial : AC a b f g → AC b c g h → AC a c f h
  parallel :
    (pa : par-qubits a1 a2 a3) →
    (pb : par-qubits b1 b2 b3) →
    (par-timing pa f1 f2 f3) →
    (par-timing pb g1 g2 g3) →
    AC a1 b1 f1 g1 →
    AC a2 b2 f2 g2 →
    AC a3 b3 f3 g3
\end{lstlisting}

The type \lstinline{constantly t f} asserts that \lstinline{f} gives time \lstinline{t} for each qubit in its domain, and the relation $\le$ on timing functions is defined pointwise.
For the \lstinline{gate} constructor it is noted that the uses of \lstinline{constant} mean the gate uses all its qubits over the same time interval; graphically, its schedule is a rectangle, that is to say its left and right edges are flat rather than ragged.

The type \lstinline{par-timing} ensures that timing functions agree for each qubit occurrence across the bijection specified by \lstinline{par-qubits}, motivated both practically by the encoding of timing functions, and also by the interest in distinguishing occurrences of qubits as described above in relation to the definition of \lstinline{par-qubits}.

Finally, annotations can be erased: each constructor of \lstinline{AC} is simply replaced by its counterpart in \lstinline{Circuit} (with \lstinline{delay} becoming \lstinline{wire}).
This is carried out by the function \lstinline{erase-timing}, with the following type:
\begin{lstlisting}[caption={erase},float=h]
erase : AC a b f g → Circuit a b
\end{lstlisting}

\subsection{As-late-as-possible scheduling}\label{sec:scheduling}

The \emph{as-late-as-possible} scheduling algorithm is implemented through the function \lstinline{alap}:
\begin{lstlisting}[caption={As Late As Possible}, float=h]
alap : (g : Timing b) → Circuit a b → Σ (Timing a) (λ f → AC a b f g)
\end{lstlisting}
It takes a right-timing \(g\) and plain circuit as arguments, and returns both an annotated circuit and its left-timing \(f\).
It is implemented by structural induction on the circuit, packing the schedule as far to the right as possible and propagating timings from right to left.

The non-trivial case is that of a \lstinline{gate}: gates operate on all of their qubits over the same time interval, hence their constant left and right timing functions in \lstinline{AC}, but the provided right timing \(g\) is arbitrary (graphically: the gate is a rectangle, while \(g\) might be jagged).
This is solved by first ``flattening'' \(g\) by inserting a delay with right-timing \(g\) and constant left-timing \(h=\text{const}(\text{min}(g))\).
The gate is then scheduled left of this delay, i.e. with right-timing \(h\).
It is observed that the \lstinline{delay} will have duration zero for at least one qubit occurrence (where \(g\) attains its minimum), so we could not have hoped to schedule the gate later, and \lstinline{alap} is indeed performing as-late-as-possible scheduling as claimed.

While the full definitions of equivalence on \lstinline{Circuit} and \lstinline{AC} are not yet complete, a rudimentary notion of equivalence, written \lstinline{_≈_}, is defined on \lstinline{Circuit}: it is the smallest relation that is reflexive, respected by \lstinline{serial} and \lstinline{parallel}, and equates \lstinline{serial ckt wire} with \lstinline{ckt}.
This form of equivalence is sufficient to show that \lstinline{alap} (with some decoration) is a valid scheduling algorithm, which is to say a right inverse to erasure:
\begin{lstlisting}[caption={erase-alap},float=h]
erase-alap : (g : Timing b) (ckt : Circuit a b) →
  (erase ∘ proj₂ ∘ alap g) ckt ≈ ckt
\end{lstlisting}
From the result at the end of Section \ref{sec:hwmodelcor} we then know that a pulse schedule generated from the annotated output of \lstinline{alap} implements the same unitary as the original circuit.

\section{Conclusion}

We have given two linear type theories that can represent quantum circuits with timing information. We have given a model in terms of the input signals to a quantum chip, in such a way that the interpretation function is the algorithm that perform compilation. We have also proved that the term models are the initial models, giving confidence that the syntax and semantics are in close correspondence.

For future work, we want to expand our language GRAMPUS with the other features used in quantum algorithms, in particular measurement, and the ability to perform hybrid algorithms where the result of a measurement affects which gates are subsequently performed.

\bibliography{refs}

\section{Appendix --- Syntactic Models}
\label{appendix:synmod}
Here we explore more formally some of the notions introduced above. We start by explicitly constructing the initial models $\text{Circ}_u$ and $\text{Circ}$ (the latter being a more explicit version of Definition \ref{def:circ}).

\begin{df}
The SMC $\text{Circ}_u$ is defined in the following way:
\begin{itemize}
\item Its objects are finite sequences of qubits $\mathbf{Ob}(\text{Circ}_u) := \mathbf{Q}^{*}$,
\item Its unit object is the empty sequence $I := []$,
\item The action of its tensor product on objects is given by concatenation,
\item Its morphisms are generated (under composition and tensor products) by permutations $[q_1, \dots, q_n] \xrightarrow{\sigma} [q_{\sigma(1)}, \dots, q_{\sigma(n)}]$ and gates in $\mathbf{G}$,
\item The unitors and associator are just identities, and
\item The symmetry $\sigma$ is the appropriate permutation.
\end{itemize}
\end{df}

The definition of a categorical model for the plain language given in the text (Definition \ref{def:PLAPUS-model}) is equivalent to saying that a model is an SMC $\mathcal{C}^\otimes$ together with a strong symmetric monoidal functor $\text{Circ}_u \to \mathcal{C}^\otimes$. An immediate consequence is then the following observation.

\begin{remark}\label{rmk:category-of-PLAPUS-models}
A symmetric monoidal functor $\mathcal{F} : \mathcal{C}^\otimes \to \mathcal{D}^\otimes$ between models of the plain language $(\mathcal{C}^\otimes, X_{-}, f_{-}), (\mathcal{D}^\otimes, Y_{-}, g_{-})$ preserves the choices of qubits and gates precisely when it makes the following diagram commute
\begin{center}\begin{tikzcd}
& \textup{Circ}_u \arrow[dl] \arrow[dr] & \\
\mathcal{C}^\otimes \arrow[rr] && \mathcal{D}^\otimes
\end{tikzcd}\end{center}
Hence the category of models is equivalent to the coslice category $\text{Circ}_u\downarrow \mathbf{SMC}$.
\end{remark}

It's important to notice that $\text{Circ}_u$ is \emph{not} the category of contexts for the plain theory (though it is symmetric-monoidally equivalent to it): it is the free \emph{strict} SMC on the data encoded by $\mathbf{Q}$ and $\mathbf{G}$, while the category of contexts would be the free SMC on such data. The equivalence between the two is a direct consequence of the general strictification theorem for SMCs (cft. with \cite{Maclane1963NaturalAA} and chapter XI of \cite{alma990012595570302711}). This, together with a simple calculation, also implies that the model $\mathbf{Syntax}_u$ (defined in Section \ref{sec:syntactic}) is equivalent to $\text{Circ}_u$, hence initial.

Similarly, we can give a very concise (though somewhat abstract) description of models of the annotated language.

\longversion{
\begin{prop}
The real number line $\mathbb{R}$ equipped with the usual ordering $\le$ and the monoid structure given by addition is an ordered monoid (i.e., $\le$ is a \emph{congruence} on $(\mathbb{R}, 0, +)$. This allows us to define a SMC $\mathbb{R}_{\le}$ as follows:
\begin{itemize}
\item Its objects are integers $\mathbf{Ob}(\mathbb{R}_{\le}) := \mathbb{R}$,
\item Its unit object is given by $0 \in \mathbb{R}$,
\item Its monoidal structure is given, on objects, by addition $x \otimes y := x + y$,
\item Its morphisms are given by inequalities (i.e. there is a morphism $f : x \to y$ iff $x \le y$),
\item Unitors, associator and symmetry are identities
\end{itemize}
Identities and compositions are well-defined because total orders are reflexive and transitive, and the tensor product is well-defined because $\le$ is a congruence.
\end{prop}
}{}

\begin{df}
Given a monoidal category $\mathcal{B}^\otimes = (\mathcal{B}, 1, \otimes, \dots)$, a \emph{symmetric monoidal} $\mathcal{B}^\otimes$-\emph{module} category $\mathcal{C}^{\boxtimes, \bullet} = (\mathcal{C}^\boxtimes, \bullet)$ ($\mathcal{B}^\otimes$-SMMC in the future, or just SMMC if no emphasis is put on $\mathcal{B}^\otimes$) is the data of an SMC $\mathcal{C}^\boxtimes = (\mathcal{C}, I, \boxtimes, \dots)$ together with a strong monoidal functor $\bullet : \mathcal{B}^\otimes \to \mathbf{SMC}[\mathcal{C}^\boxtimes, \mathcal{C}^\boxtimes]^\circ$ into the monoidal (under composition) category of strong symmetric monoidal endofunctors of $\mathcal{C}^\boxtimes$.
\end{df}

Just as we did for $\text{Circ}_u$, we now provide a more formal definition of $\text{Circ}$:

\begin{df}\label{def:circ-appendix}
The $\mathbb{R}$-SMMC $\text{Circ}$ is defined as follows:
\begin{itemize}
\item Its objects are finite sequences of pairs $(d, q)$ of a real number and a qubit: $\mathbf{Ob}(\text{Circ}) := (\mathbb{R} \times \mathbf{Q})^{*}$,
\item Its unit object is the empty sequence,
\item Its tensor product is given, on objects, by concatenation,
\item Its morphisms are generated (under composition and tensor products) by permutations, delays for every qubit $q$ and real number $x$
\[ \text{delay}_d : [(x, q)] \to [(x + d, q)] \]
and gates for every $(G, d, q_1, \dots, q_n) \in \mathbf{G}$
\[ \phi_G : [(x, q_1), \dots, (x, q_n)] \to [(x + d, q_1), \dots, (x + d, q_n)] \]
\item Unitors and associators are just identities,
\item Symmetries are the appropriate permutations,
\item For every real number $d \in \mathbb{R}$, the endofunctor $\bullet(d)$ acts
\begin{itemize}
\item On objects, by adding $d$ to the first component of every element of the sequence,
\[ \bullet(d)([(x_1, q_1), \dots, (x_n, q_n)]) := [(x_1 + d, q_1), \dots, (x_n + d, q_n)] \]
\item On morphisms, it is the identity on permutations, maps delays to delays appropriately, and similarly for gates.
\end{itemize}
\end{itemize}
\end{df}

In total analogy with the case for the plain language, the definition of a categorical model of the annotated language (Definition \ref{def:GRAMPUS-model}) is equivalent to saying that a model is an SMMC $\mathcal{C}^{\otimes, \bullet}$ together with an SMMC-morphism\footnote{Defined as a strong linear functor, 3.3.2 in \cite{capucci2024actegoriesworkingamthematician}, such that the underlying functor is strong symmetric monoidal and that the ``lineator'' is a strong symmetric monoidal transformation} $\text{Circ} \to \mathcal{C}^{\otimes,\bullet}$. We also notice the following fact.

\begin{remark}
An SMMC-morphism $\mathcal{C}^{\otimes, \bullet} \to \mathcal{D}^{\otimes, \bullet}$ preserves the choices of qubits, gates and delays precisely when it makes the following diagram commute
\begin{center}\begin{tikzcd}
& \textup{Circ} \arrow[dl] \arrow[dr] & \\
\mathcal{C}^{\otimes, \bullet} \arrow[rr] && \mathcal{D}^{\otimes, \bullet}
\end{tikzcd}\end{center}
Hence the category of models is equivalent to the coslice category $\text{Circ} \downarrow \mathbb{R}\text{-}\mathbf{SMMC}$.
\end{remark}

Considerations analogous to those made about $\text{Circ}_{u}$ apply here as well, though it is easier to prove the equivalence with the category of contexts directly.

\longversion{
The appropriate strictification theorem can be proven as a corollary of the universal property of the free SMMC construction, defined in total analogy with 2.1 in \cite{capucci2024actegoriesworkingamthematician}. More precisely (and abstractly), such a universal property can be seen as a special case of the pseudo morphism classifier (3.13 in \cite{BLACKWELL19891}) where the 2-category is that of monoidal categories and the 2-monad is given by $\mathcal{C}^\otimes \mapsto \mathbb{R} \times \mathcal{C}^{\otimes}$.\lp{Does it make sense to sketch the argument here, without giving details? The full proof is rather technical, and the resulting category being \emph{symmetric} monoidal, though a consequence of $\mathbb{R}$}
}{}

\section{Appendix --- Commuting Conversions}
\label{appendix:comcon}
The full list of commuting conversions is given below. Each one should be given as a rule of deduction with premises that ensure the left-hand side is well-typed (which ensures the right-hand side is well-typed too). We omit these to save space here.

\begin{align*}
(\text{let}\ * = s \text{ in let}\ * = t \text{ in } u)
& = (\text{let}\ * = \text{let}\ * = s \text{ in } t \text{ in } u) \\
(\text{let}\ * = s \text{ in let}\ * = t \text{ in } u)
& = (\text{let}\ * = t \text{ in let}\ * = s \text{ in } u) \\
(\text{let} \ * = s \text{ in } G(t_1, \ldots, t_n)) & = G(t_1, \ldots, t_{i-1},\\
& \qquad \text{let}\ * = s \text{ in } t_i, \\
& \qquad t_{i+1}, \ldots, t_n) \\
(\text{let}\ * = s \text{ in } (t,u)) & = (\text{let}\ * = s \text{ in } t, u) \\
(\text{let}\ * = s \text{ in } (t,u)) & = (t, \text{let}\ * = s \text{ in } u) \\
(\text{let}\ * = s \text{ in let}\ (x,y) = t \text{ in } u)
& = (\text{let}\ (x,y) = \text{let}\ * = s \text{ in } t \text{ in } u)
\end{align*}

\begin{align*}
(\text{let}\ * = s \text{ in let}\ (x,y) = t \text{ in } u)
& = (\text{let}\ (x,y) = t \text{ in let}\ * = s \text{ in } u) \\
(\text{let}\ * = s \text{ in box}\ d\ t) & = \text{box}\ d (\text{let}\ * = s \text{ in } t) \\
(\text{let}\ * = s \text{ in let box}\ d\ x = t \text{ in } u)
& = (\text{let box}\ d\ x = \text{let}\ * = s \text{ in } t \text{ in } u) \\
(\text{let}\ * = s \text{ in let box}\ d\ x = t \text{ in } u)
& = (\text{let box}\ d\ x = t \text{ in let}\ * = s \text{ in } u) \\
(\text{let}\ (x,y) = s \text{ in let}\ * = t \text{ in } u)
& = (\text{let}\ * = \text{let}\ (x,y) = s \text{ in } t \text{ in } u) \\
(\text{let}\ (x,y) = s \text{ in let}\ * = t \text{ in } u)
& = (\text{let}\ * = t \text{ in let}\ (x,y) = s \text{ in } u) \\
(\text{let} \ (x,y) = s \text{ in } G(t_1, \ldots, t_n)) & = G(t_1, \ldots, t_{i-1}, \\
& \qquad \text{let}\ (x,y) = s \text{ in } t_i, \\
& \qquad t_{i+1}, \ldots, t_n)
\end{align*}

\begin{align*}
(\text{let}\ (x,y) = s \text{ in } (t,u)) & = (\text{let}\ (x,y) = s \text{ in } t, u) \\
(\text{let}\ (x,y) = s \text{ in } (t,u)) & = (t, \text{let}\ (x,y) = s \text{ in } u) \\
(\text{let}\ (x,y) = s \text{ in let}\ (z,w) = t \text{ in } u)
& = (\text{let}\ (z,w) = \text{let}\ (x,y) = s \text{ in } t \text{ in } u) \\
(\text{let}\ (x,y) = s \text{ in let}\ (z,w) = t \text{ in } u)
& = (\text{let}\ (z,w) = t \text{ in let}\ (x,y) = s \text{ in } u) \\
(\text{let}\ (x,y) = s \text{ in box}\ d\ t) & = \text{box}\ d (\text{let}\ (x,y) = s \text{ in } t) \\
(\text{let}\ (x,y) = s \text{ in let box}\ d\ z = t \text{ in } u)
& = (\text{let box}\ d\ z = \text{let}\ (x,y) = s \text{ in } t \text{ in } u) \\
(\text{let}\ (x,y) = s \text{ in let box}\ d\ z = t \text{ in } u)
& = (\text{let box}\ d\ z = t \text{ in let}\ (x,y) = s \text{ in } u) \\
(\text{let}\ \text{box}\ d\ z = s \text{ in let}\ * = t \text{ in } u)
& = (\text{let}\ * = \text{let}\ \text{box}\ d\ z = s \text{ in } t \text{ in } u)
\end{align*}

\begin{align*}
(\text{let}\ \text{box}\ d\ z = s \text{ in let}\ * = t \text{ in } u)
& = (\text{let}\ * = t \text{ in let}\ \text{box}\ d\ z = s \text{ in } u) \\
(\text{let} \ \text{box}\ d\ z = s \text{ in } G(t_1, \ldots, t_n)) & = G(t_1, \ldots, t_{i-1}, \\
& \qquad \text{let}\ \text{box}\ d\ z = s \text{ in } t_i, \\
& \qquad t_{i+1}, \ldots, t_n) \\
(\text{let}\ \text{box}\ d\ z = s \text{ in } (t,u)) & = (\text{let}\ \text{box}\ d\ z = s \text{ in } t, u) \\
(\text{let}\ \text{box}\ d\ z = s \text{ in } (t,u)) & = (t, \text{let}\ \text{box}\ d\ z = s \text{ in } u) \\
(\text{let}\ \text{box}\ d\ z = s \text{ in let}\ (x,y) = t \text{ in } u)
& = (\text{let}\ (x,y) = \text{let}\ \text{box}\ d\ z = s \text{ in } t \text{ in } u) \\
(\text{let}\ \text{box}\ d\ z = s \text{ in let}\ (x,y) = t \text{ in } u)
& = (\text{let}\ (x,y) = t \text{ in let}\ \text{box}\ d\ z = s \text{ in } u) \\
(\text{let}\ \text{box}\ d\ z = s \text{ in box}\ d'\ t) & = \text{box}\ d'\ (\text{let}\ \text{box}\ d\ z = s \text{ in } t) \\
(\text{let}\ \text{box}\ d'\ z = s \text{ in let box}\ d\ x = t \text{ in } u)
& = (\text{let box}\ d\ x = t \text{ in let}\ \text{box}\ d'\ z = s \text{ in } u)
\end{align*}
\end{document}